\documentclass[12pt]{iopart}
\usepackage{graphicx}
\usepackage{hyperref}
\usepackage{iopams}

\begin{document}

\article[Spin squeezing in cold ${\it ^{87}}$Rb]{Special
Issue/ST}{Conditions for spin squeezing in a cold $\mathbf{^{87}}$Rb
ensemble}

\author{S R de Echaniz$^{\dag}$, M W Mitchell$^{\dag}$, M
Kubasik$^{\dag}$, M Koschorreck$^{\dag}$, H Crepaz$^{\dag}$, J
Eschner$^{\dag}$  and E S Polzik$^{\ddag}$}

\address{\dag\ ICFO - Institut de Ci\`{e}ncies Fot\`{o}niques, E-08034 Barcelona, Spain}
\address{\ddag\ QUANTOP, Niels Bohr Institute, Copenhagen
University, Blegdamsvej 17, DK-2100 K{\o}benhavn, Denmark}

\ead{sebastian.echaniz@icfo.es}

\begin{abstract}
We study the conditions for generating spin squeezing via a quantum
non-demolition measurement in an ensemble of cold $^{87}$Rb atoms.
By considering the interaction of atoms in the $5S_{1/2}(F=1)$
ground state with probe light tuned near the D$_2$ transition, we
show that, for large detunings, this system is equivalent to a
spin-1/2 system when suitable Zeeman substates and quantum operators
are used to define a pseudo-spin. The degree of squeezing is derived
for the rubidium system in the presence of scattering causing
decoherence and loss. We describe how the system can decohere and
lose atoms, and predict as much as 75\% spin squeezing for atomic
densities typical of optical dipole traps.
\end{abstract}


\submitto{\JOB}


\section{Introduction}

There has recently been much interest in coupling light with atomic
ensembles to develop a quantum interface. Several proposals have
been published to utilise this kind of coupling for spin squeezing
\cite{Kuzmich1997, Kuzmich1998, Thomsen2002}, quantum memories
\cite{Kozhekin2000}, quantum teleportation \cite{Kuzmich2000},
entanglement \cite{Duan2000}, magnetometry \cite{Geremia2003} and
atomic clocks \cite{Oblak2005}. Many of these proposals have been
realised experimentally using samples of alkali atoms in vapour
cells and in magneto-optical traps (MOT) \cite{Hald1999,
Kuzmich2000a, Schori2001, Julsgaard2004, Julsgaard2001,
Geremia2004}. Spin squeezing is the simplest of these applications,
and is often regarded as a benchmark of the light-atomic-ensemble
interaction. It has been demonstrated a few times: first in a MOT by
mapping squeezed states of light onto the atomic spin
\cite{Hald1999}, then in vapour cells via a quantum non-demolition
(QND) measurement \cite{Kuzmich2000a}, and recently using the same
method in a MOT but with the help of feedback \cite{Geremia2004}.

In this article we study the conditions for generating spin
squeezing via a QND measurement \cite{Kuzmich1998} in a cold
ensemble of $^{87}$Rb atoms using the $5S_{1/2}(F=1)$ ground state
of the D$_2$ transition. We show that this system is formally
equivalent to a spin-1/2 system when suitable Zeeman substates and
quantum operators are used to define a pseudo-spin. In this scheme,
we expect to have a higher light-atomic-ensemble coupling than in
previous work even when possible sources of decoherence and loss
arising from these choices are identified and taken into account.

This article is organised into five sections. The next and second
section describes the interaction between atoms and light considered
here. \Sref{sec:Rbtohalf} shows how the complicated $^{87}$Rb system
can be reduced to an effective spin-1/2 system. In
\sref{sec:DegSqueez}, we calculate the degree of squeezing
attainable in the presence of decoherence and loss. Finally, we
present the conclusions in \sref{sec:Conc}.

\section{Spin-squeezing interaction}

\begin{figure}
    \begin{center}
        \includegraphics[width=15cm]{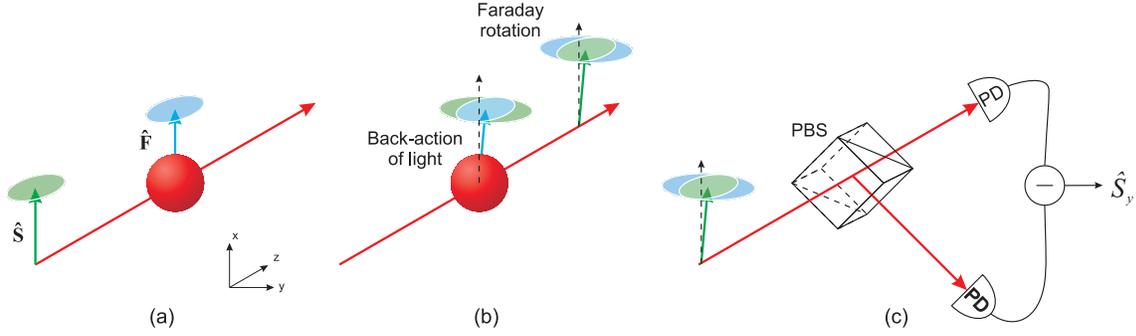}
    \end{center}
    \caption{Spin squeezing interaction: (a)  preparation of the
    initial coherent states, (b) light-atom dipole interaction and
    entanglement of quantum fluctuations, and (c) polarimetric
    measurement of the probe light using a polarising beamsplitter
    (PBS) and a couple of photodetectors (PD).}
    \label{fig:SpinInteraction}
\end{figure}

Spin squeezing can be created by using a polarised off-resonant
pulse of light to perform a QND measurement of the spin
\cite{Kuzmich1998}. First, the Stokes vector $\mathbf{\hat{S}}$
(polarisation) of the probe pulse and the spin vector
$\mathbf{\hat{F}}$ of the atomic ensemble are prepared in a coherent
state pointing in the $x$-direction (\fref{fig:SpinInteraction}(a)).
As we send the pulse through the sample, the light and atoms
interact via the dipole interaction, which in this kind of schemes
is described by the Hamiltonian
\begin{equation}
    \label{eq:Hamiltonian}
    \hat H_{SS}  = \hbar \Omega \hat S_z \hat F_z ,
\end{equation}
where $\Omega$ is a coupling strength, $\hat S_z$ is the $z$
component of the Stokes vector $\mathbf{\hat S}$ of light, and $\hat
F_z$ is the $z$ component of the atomic spin vector $\mathbf{\hat
F}$. In this interaction, the polarisation of the light is rotated
due to the Faraday effect and there is a back action of the light
onto atoms which rotates the orientation of the spin
(\fref{fig:SpinInteraction}(b)), and at the same time, their quantum
fluctuations become entangled \cite{Kupriyanov2005}. If this
interaction acts for a time $\tau$, then for small $\Omega \tau$, it
produces the following relation between the fluctuations of
$\mathbf{\hat S}$ and $\mathbf{\hat F}$ \cite{Book}:
\begin{equation}
    \label{eq:QND}
    \eqalign{
    \delta \hat F_y^{out} &= \delta \hat F_y^{in} + \Omega \tau \left\langle \hat{F}_x \right\rangle \delta \hat S_z^{in},\\
    \delta \hat F_z^{out} &= \delta \hat F_z^{in}, \\
    \delta \hat S_y^{out} &= \delta \hat S_y^{in} + \Omega \tau \left\langle \hat{S}_x \right\rangle \delta \hat F_z^{in},\\
    \delta \hat S_z^{out} &= \delta \hat S_z^{in}.}
\end{equation}
As can be seen, a measurement of $\hat{S}_y^{out}$ with a
polarimeter (\fref{fig:SpinInteraction}(c)) contains information
about the spin component $\hat{F}_z$. This QND measurement leads to
squeezing of the fluctuations $\delta \hat{F}_z$. If we ignore
decoherence and loss mechanisms, the degree of squeezing has been
shown \cite{Duan2000, Geremia2005, Hammerer2004, Madsen2004} to be
\begin{equation}
    \label{eq:xirhoeta}
    \xi ^2  = \frac{1}{{1 + \rho_0 \eta }},
\end{equation}
with $\rho_0$ being the resonant optical density and $\eta$ the
integrated spontaneous emission rate (number of photons scattered
per atom over a probe pulse). The degree of squeezing is defined
such that $\xi^2=1$ for a coherent state and $\xi^2<1$ for a
squeezed state.

\section{Reduction of the $^{87}$Rb system to an effective spin-1/2 system
    \label{sec:Rbtohalf}}

\begin{figure}
    \begin{center}
        \includegraphics[width=15cm]{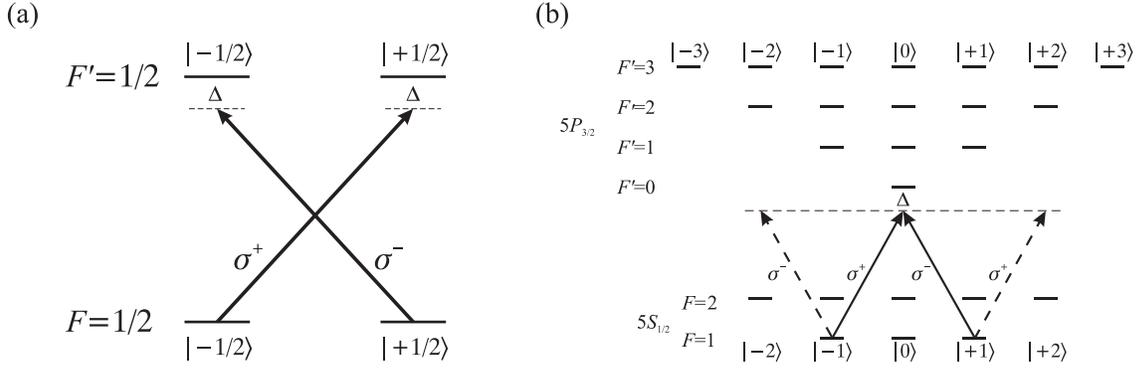}
        \caption{QND interaction between the atomic spin and the light
        polarisation in (a) an ideal spin-1/2 system, and (b) in the
        $^{87}$Rb system (solid arrows show the effective spin-1/2 system).}
        \label{fig:IdealRbSystems}
    \end{center}
\end{figure}

The ideal case of a spin-$1/2$ system as the one depicted in
\fref{fig:IdealRbSystems}(a) is simple to consider
\cite{Kuzmich1998}. In this system, the $\sigma^{+}$ and
$\sigma^{-}$ modes of the field interact with four-level atoms of
spin 1/2. After adiabatically eliminating the excited states, this
interaction is described by an interaction Hamiltonian of the form
(\ref{eq:Hamiltonian}), resulting in the typical relations
(\ref{eq:QND}) between $\mathbf{\hat{S}}$ and $\mathbf{\hat{F}}$.
Finally, we can squeeze the atomic spin by performing a QND
measurement through a measurement of $\hat{S}_{y}^{(out)}$.

When realising this kind of interaction in a realistic system like
Rb, one has to consider a more complicated, high-spin-number system,
and therefore reformulate the problem. In our particular case of
$^{87}$Rb, the lowest spin number is 1 (see
\fref{fig:IdealRbSystems}(b)).

One possible realisation in $^{87}$Rb is to use a coherent
superposition of the $|5S_{1/2}, \; F=1, \; m_F=-1\rangle$ and
$|5S_{1/2}, \; F=1, \; m_F=+1\rangle$ levels ($|-\rangle$ and
$|+\rangle$ from now on) as shown in \fref{fig:IdealRbSystems}(b).
In this case, the chosen quantum observables are components of the
alignment tensor, namely $\hat{T}_x = \hat{F}_{x}^{2} -
\hat{F}_{y}^{2}$ and $\hat{T}_y = \hat{F}_{x} \hat{F}_{y} +
\hat{F}_{y} \hat{F}_{x}$, and $\hat{F}_{z}$ \cite{Hald2001}. In fact
\begin{eqnarray}
    \eqalign{
    \hat T_x &= \left| - \right\rangle \left\langle + \right| + \left| + \right\rangle \left\langle - \right|, \\
    \hat T_y &= i \left( {\left| - \right\rangle \left\langle + \right| - \left| + \right\rangle \left\langle - \right|} \right), \\
    \hat F_z &= \left| + \right\rangle \left\langle + \right| - \left| - \right\rangle \left\langle - \right|.}
\end{eqnarray}

We can now define a collective pseudo-spin $\mathbf{\hat{J}}$ by
\begin{equation}
    \eqalign{
    \hat J_x &\equiv \frac{1}{2}\sum\limits_k {\hat T_x^k } , \\
    \hat J_y &\equiv \frac{1}{2}\sum\limits_k {\hat T_y^k } , \\
    \hat J_z &\equiv \frac{1}{2}\sum\limits_k {\hat F_z^k } ,}
\end{equation}
where the superscript $k$ denotes the single-atom operators and we
sum over all atoms. This definition fulfils the angular-momentum
commutation relations
\begin{equation}
    \left[ {\hat J_i ,\,\hat{J}_j} \right] = i \epsilon_{ijk} \hat{J}_k,
\end{equation}
when $F=1$, where $\epsilon_{ijk}$ is the Levi-Civita tensor. Hence,
one could squeeze the pseudo-spin $\mathbf{\hat{J}}$ along the
$z$-axis, as in the ideal case, if a QND-type interaction
(\ref{eq:Hamiltonian}) exists between $\mathbf{\hat{J}}$ and
$\mathbf{\hat{S}}$.

This interaction can be derived as follows. Consider the dipole
interaction Hamiltonian for an off-resonant field
\cite{Geremia2005,Happer1972}
\begin{equation}
    \hat H_{int} = \sum\limits_{F,F'} {{\bf \hat E}^{( - )} \cdot \frac{{\boldsymbol{ \hat \alpha }_{F,F'}}}{{\hbar \Delta_{F,F'}}} \cdot {\bf \hat E}^{( + )}},
\end{equation}
where ${\bf \hat E}^{( \pm )}$ are the positive and negative
frequency field operators of the probe field, $\Delta_{F,F'}$ is the
detuning of the probe from the $F \rightarrow F'$ transition of the
D$_2$ line in $^{87}$Rb, and $\boldsymbol{ \hat \alpha }_{F,F'}$ is
the atomic polarisability tensor of the transition. The latter is a
rank-2 spherical tensor, which can be decomposed into the direct sum
of a scalar, a vector and a tensor term: $\boldsymbol{ \hat \alpha
}_{F,F'} = \boldsymbol{ \hat \alpha }^{(0)}_{F,F'} \oplus
\boldsymbol{ \hat \alpha }^{(1)}_{F,F'} \oplus \boldsymbol{ \hat
\alpha }^{(2)}_{F,F'}$. This leads to the decomposition of the
interaction Hamiltonian into $\hat H_{int}  = \hat H^{(0)}  + \hat
H^{(1)} + \hat H^{(2)}$, that can be expressed as \cite{Geremia2005}
\numparts
\begin{eqnarray}
    \label{eq:H0}
    \hat H^{(0)} & = \alpha _0 g\sum\limits_{F'} {\frac{{\alpha _{F,F'}^{(0)} }}{{\Delta _{F,F'} }}} \hat n\hat N, \\
    \label{eq:H1}
    \hat H^{(1)} & = \alpha _0 g\sum\limits_{F'} {\frac{{\alpha _{F,F'}^{(1)} }}{{\Delta _{F,F'} }}} \hat S_z \hat J_z , \\
    \label{eq:H2}
    \hat H^{(2)} & = \alpha _0 g\sum\limits_{F'} {\frac{{\alpha _{F,F'}^{(2)} }}{{\Delta _{F,F'} }}} \left[ {\hat S_x \hat J_x  - \hat S_y \hat J_y  + \frac{{2\hat n \hat N}}{{\sqrt 6 }}} \right],
\end{eqnarray}
\endnumparts
when we consider the transitions from the ground states $|F=1, \;
m_F = \pm 1 \rangle$ to all the excited states $F'$. Here $\alpha
_{F,F'}^{(i)}$ is the rank-$i$ unitless polarisability coefficient
of the $F \rightarrow F'$ transition, $\hat{n}$ ($\hat{N}$) is the
photon (atom) number operator, and
\begin{eqnarray}
    \eqalign{
    \alpha _0  &= \frac{{3\epsilon _0 \hbar \Gamma \lambda _0^3 }}{{8\pi ^2 }}, \\
    g &= \frac{{\omega _0 }}{{2\epsilon _0 V}} ,}
\end{eqnarray}
with $\Gamma$ being the spontaneous decay rate, $\lambda_0$
($\omega_0$) the transition wavelength (frequency), and $V$ the
interaction volume. The components of the Stokes vector can be
expressed in terms of the annihilation and creation operators of the
$\sigma^{\pm}$ modes of the field
\begin{equation}
    \eqalign{
    \hat S_x &= \frac{1}{2}\left( {\hat a_-^{\rm \dag } \hat a_+ + \hat a_+^{\rm \dag } \hat a_-} \right), \\
    \hat S_y &= \frac{i}{2}\left( {\hat a_-^{\rm \dag } \hat a_+ - \hat a_+^{\rm \dag } \hat a_-} \right), \\
    \hat S_z &= \frac{1}{2}\left( {\hat a_+^{\rm \dag } \hat a_+ - \hat a_-^{\rm \dag } \hat a_-} \right),}
\end{equation}
which can be related to the field operators ${\bf \hat E}^{( \pm
)}$.

As can be seen from (\ref{eq:H0}), $\hat H^{(0)}$ is proportional to
$\hat n \hat N$ and can be viewed as a global phase shift common to
both polarisation modes of the probe pulse, which does not produce
any signal on the output of the polarimeter, and therefore can be
omitted. The term $\hat{S}_x \hat{J}_x - \hat{S_y} \hat{J}_y$ in the
rank-2 Hamiltonian (\ref{eq:H2}) corresponds to Raman transitions
between $\left| + \right\rangle$ and $\left| - \right\rangle$, the
other term is also proportional to $\hat n \hat N$, and can be
omitted. Furthermore, for the case of detunings much larger than the
hyperfine splitting of the excited states ($|\Delta_{F,F'}| \gg
|\Delta_{hfs}^e|$), the sum over $\alpha _{F,F'}^{(2)}$ tends to
zero, and hence $\hat H^{(2)}$ can be neglected. A similar situation
occurs for $\hat H^{(1)}$ in (\ref{eq:H1}), but this time only the
contributions from $F'=1,2$ tend to zero when $|\Delta_{F,F'}| \gg
|\Delta_{hfs}^e|$, leaving just those from $F'=0$.

Altogether, we are left with an effective 3-level $\Lambda$ system
formed by the $\left| \pm \right\rangle$ ground states and the
$F'=0$ excited state (solid arrows in \fref{fig:IdealRbSystems}(b)),
which doesn't exhibit Raman transitions, and therefore is formally
equivalent to the simple spin-1/2 system of
\fref{fig:IdealRbSystems}(a). The remaining effective Hamiltonian is
\begin{equation}
    \hat H_{eff}  = \alpha _0 g\frac{{\alpha _{1,0}^{(1)} }}{\Delta_{1,0}}\hat S_z \hat J_z ,
\end{equation}
which is of the form (\ref{eq:Hamiltonian}) necessary for the QND
interaction.

\section{Degree of squeezing in the presence of scattering
    \label{sec:DegSqueez}}

We now consider the degree of squeezing for a general spin-$F$
system. The degree of squeezing defined by Wineland \etal
\cite{Wineland1992} for a frequency standard based on the Ramsey
method is
\begin{equation}
    \label{eq:xi}
    \xi ^2  \equiv \frac{{\left\langle {\left( {\Delta \hat F_z } \right)^2 } \right\rangle }}{{\left| {\left\langle {\hat F_x } \right\rangle } \right|^2 }} 2 N F,
\end{equation}
which implies entanglement between the individual atomic spins when
there is squeezing \cite{Haldthesis}.

In the presence of scattering, atoms can be lost when they are
pumped out of the initial atomic system, or can undergo decoherence
when they stay within the system. These two cases are discussed in
the following subsections, where we denote by $\beta$ the number of
scattered photons which produce loss, and by $\gamma$ those that
produce decoherence, with $\eta = \beta + \gamma$.

\subsection{Atom loss and decoherence}

The case of atom loss occurs when the atoms are pumped out of the
initial atomic system due to e.g. collisions with the background or
spontaneous decay into a state not participating in the interaction.

In this case, it can be shown \cite{MolmerPriv} that the variance
$\langle (\Delta \hat F'_z)^2 \rangle$ of the remaining
$N'=(1-\beta)N$ atoms is
\begin{equation}
    \label{eq:lostvar}
    \left\langle {\left( {\Delta \hat F'_z } \right)^2 } \right\rangle  = \left( {1 - \beta } \right)^2 \left\langle {\left( {\Delta \hat F_z } \right)^2 } \right\rangle  + \beta \left( {1 - \beta } \right)N\frac{F}{2}.
\end{equation}
Using equation \eref{eq:xi}, the degree of squeezing for the
remaining atoms is
\begin{equation}
    \label{eq:xilost}
    \xi '^2  = \left( {1 - \beta } \right)\xi ^2  + \beta .
\end{equation}

We now assume that $\gamma N$ atoms undergo decoherence due to
scattering within the initial coherent superposition. The total
variance of $\hat F'_z$ in this case is transformed in the same way
as (\ref{eq:lostvar}), but with the added variance
$\mathrm{var}(\hat F'_z)_\gamma$ of the individual decohered atoms:
\begin{equation}
    \label{eq:decovar}
    \left\langle {\left( {\Delta \hat F'_z } \right)^2 } \right\rangle = \left( {1 - \gamma } \right)^2 \left\langle {\left( {\Delta \hat F_z } \right)^2 } \right\rangle + \gamma \left( 1-\gamma \right) N \frac{F}{2} + \gamma N \mathrm{var}(\hat F'_z)_\gamma,
\end{equation}
and the degree of squeezing will be
\begin{equation}
    \label{eq:xideco}
    \xi '^2  = \xi ^2 + \frac{\gamma}{1 - \gamma} + \frac{2 \mathrm{var}(\hat F'_z)_\gamma}{F} \frac{\gamma}{\left( 1-\gamma \right) ^2}.
\end{equation}

\subsection{Spin-1/2 and $^{\it 87}$Rb systems}

\begin{figure}
    \begin{center}
        \includegraphics[width=7cm]{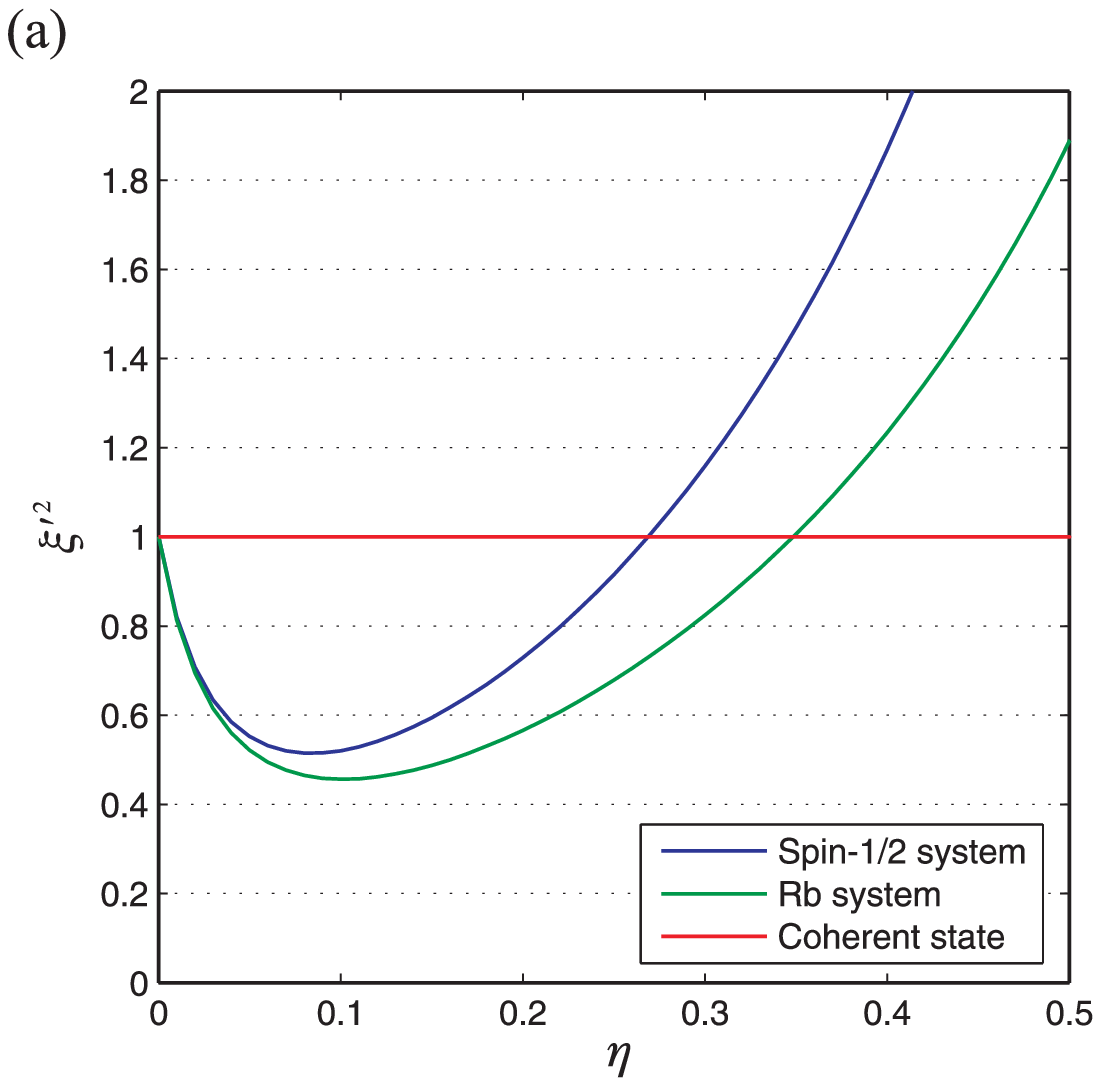}
        \includegraphics[width=7cm]{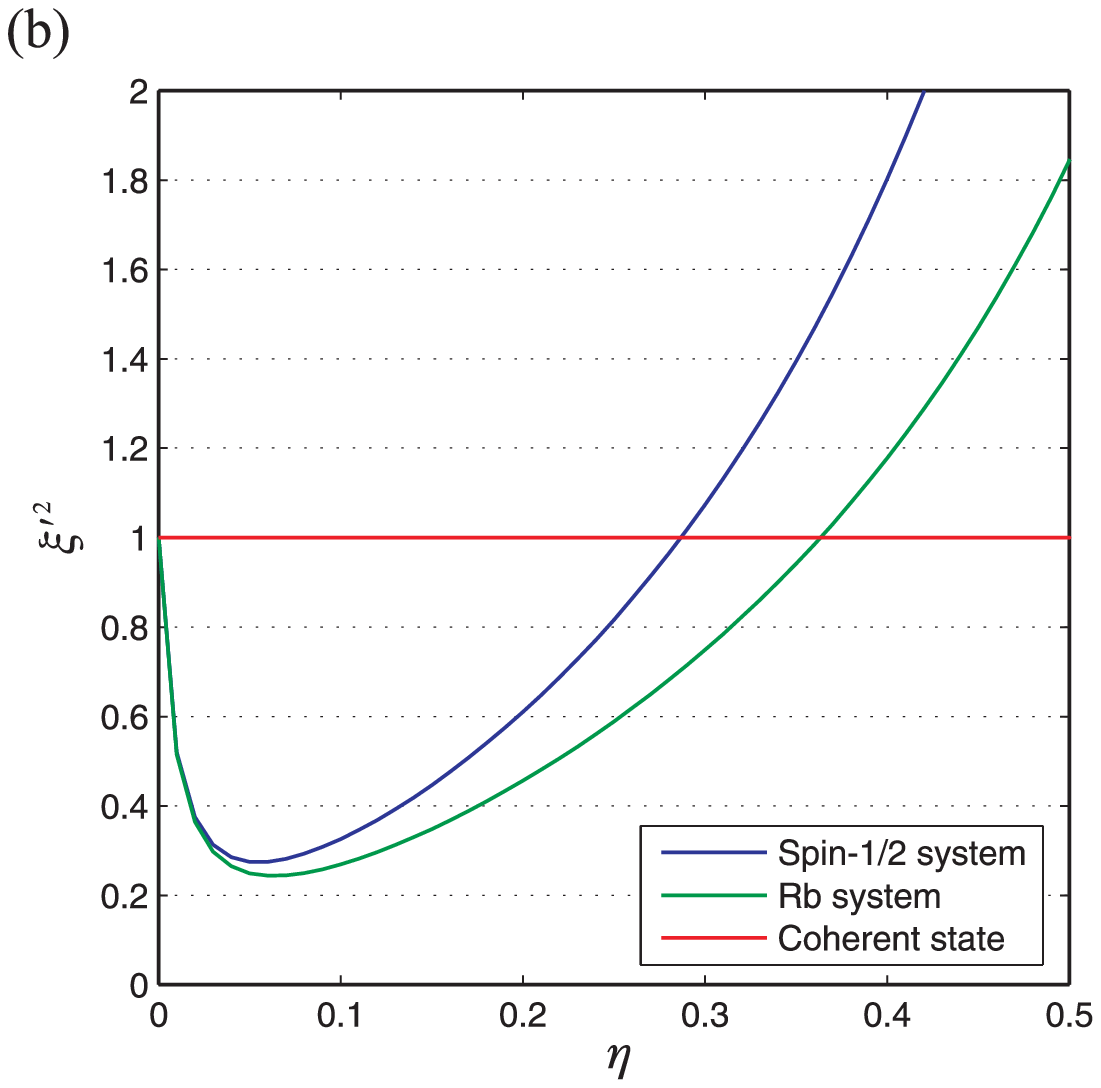}
        \caption{Degree of squeezing $\xi'^2$ as a function of the
        integrated scattering rate $\eta$ for (a) a standard MOT
        ($\rho_0=25$) and (b) a standard FORT ($\rho_0=100$), and for
        a coherent state (red curve), the spin-1/2 system (blue curve)
        and the $^{87}$Rb system (green curve).}
        \label{fig:degsqueez}
    \end{center}
\end{figure}

If we now consider the spin-1/2 system of
\fref{fig:IdealRbSystems}(a), we notice that in this scheme, atoms
can only undergo decoherence due to photon scattering ($\gamma =
\eta, \; \beta = 0$). Hence, the degree of squeezing can be promptly
calculated from \eref{eq:xirhoeta} and \eref{eq:xideco}, and taking
into account that the variance of each of the decohered atoms is
$\mathrm{var}(\hat F'_z)_\gamma = 1/4$. \Fref{fig:degsqueez} shows
the calculated degree of squeezing as a function of the integrated
scattering rate (blue curves) for (a) a standard MOT with
$\rho_0=25$ and (b) a typical far off-resonant trap (FORT) with
$\rho_0=100$.

The case of $^{87}$Rb is again more complicated. Assuming the probe
light interacts only with the $|F=1\rangle$ ground state, i.e.\
$|\alpha_{1,F'}^{(i)} / \Delta _{1,F'}| \gg |\alpha_{2,F'}^{(i)} /
\Delta _{2,F'}|$, if the atoms decay into the $\left| F=2
\right\rangle$ hyperfine ground state, they do not interact with the
light any more. Furthermore, if they decay in the $\left| F=1, m_F=0
\right\rangle$ Zeeman substate, they interact with both polarisation
modes equally (see \fref{fig:IdealRbSystems}(b)) and do not produce
any signal at the polarimeter. Hence all these atoms have fallen out
of the considered pseudo-spin system, and from the perspective of
the light-atom interaction, these atoms are simply lost. On the
other hand, if the atoms are excited and decay back to the relevant
states $| \pm \rangle$ they will reduce the coherence of the system.

Generalising the expressions above for the pseudo-spin
$\mathbf{\hat{J}}$ and putting \eref{eq:xirhoeta}, \eref{eq:xilost}
and \eref{eq:xideco} together, we can arrive to the following
expression for the degree of squeezing in the presence of
decoherence and loss
\begin{equation}
    \xi '^2  = \frac{1 - \beta }{1 + \rho_0\eta} + \eta \frac {1 - \beta}{1 - \eta} + \gamma \frac{1 - \beta}{\left( 1 - \eta \right)^2},
\end{equation}
where we have used $\mathrm{var}(\hat J'_z)_\gamma = 1/4$. In our
$^{87}$Rb system $\gamma = \frac{5}{3} \beta$, according to the
branching ratios that determine how $\eta$ splits. This expression
is shown as green curves on \fref{fig:degsqueez}. Notice that
although this system is an effective spin-1/2 system, it outperforms
the ideal spin-1/2 system, illustrating the fact that spin squeezing
is more robust against loss than against decoherence, as can be seen
by comparing equations (\ref{eq:lostvar}) and (\ref{eq:decovar}).

As it is shown on \fref{fig:degsqueez}, for a given value of
$\rho_0$ there is an optimum value of $\eta$ for which $\xi'^2$ is
minimum. This arises from the fact that although scattering produces
decoherence and loss, a certain degree of scattering is needed for
the atoms to interact with the probe light. As much as 55\%
squeezing can be achieved for $\eta = 0.10$ in a standard MOT
($\rho_0 = 25$) and 75\% for $\eta = 0.06$ in a typical FORT
($\rho_0 = 100$).

\section{Conclusion}
    \label{sec:Conc}

We have presented a scheme in $^{87}$Rb to perform spin squeezing of
an atomic ensemble via a QND measurement and compared it to an ideal
spin-1/2 system. We have found that the rubidium system can be
reduced to an effective spin-1/2 system for large detunings
($|\Delta_{F,F'}| \gg |\Delta_{hfs}^e|$) by considering the
different tensor components of the atomic polarisability and
choosing suitable optical polarisation and atomic states, and with
the help of a pseudo-spin defined in terms of the alignment tensor.

The degree of squeezing is derived for the rubidium system in the
presence of scattering causing decoherence and loss, showing that it
is more robust to loss than to decoherence. We describe how the
system can decohere and lose atoms, and identify a minimum on
$\xi'^2$ for a given value of $\rho_0$, which arises from the
competition between the destructive scattering of photons and the
desirable coupling between light and atoms. As much as 75\%
squeezing at $\eta = 0.06$ is predicted for a typical FORT with
$\rho_0 = 100$.

\ack

This work is funded by the Spanish Ministry of Science and Education
under the LACSMY project (Ref. FIS2004-05830).


\end{document}